\documentclass[preprint,superscriptaddress]{revtex4-1}
\usepackage{CJK}

\usepackage{amsmath}
\usepackage{graphicx}
\usepackage{extarrows}
\usepackage[caption=false]{subfig}

\begin{document}
\begin{CJK*}{GB}{} 


\title{Dramatic increase in the thermal boundary conductance and radiation limit from a Nonequilibrium Landauer Approach}



\author{Jingjing Shi}
\affiliation{School of Mechanical Engineering and Birck Nanotechnology Center, Purdue University, West Lafayette, Indiana 47907, USA}

\author{Xiaolong Yang}
\affiliation{School of Mechanical Engineering and Birck Nanotechnology Center, Purdue University, West Lafayette, Indiana 47907, USA}
\affiliation{Frontier Institute of Science and Technology, Xi'an Jiaotong University, Xi'an 710049, People's Republic of China}

\author{Timothy S. Fisher}
\affiliation{Department of Mechanical and Aerospace Engineering \& California NanoSystems Institute, University of California Los Angeles, Los Angeles, CA 90095, USA}

\author{Xiulin Ruan}
\email{ruan@purdue.edu}
\affiliation{School of Mechanical Engineering and Birck Nanotechnology Center, Purdue University, West Lafayette, Indiana 47907, USA}


\date{\today}

\begin{abstract}
Thermal boundary conductance (TBC) is critical in many thermal and energy applications. A decades-old puzzle has been that many of the measured TBCs, such as those well characterized across Al/Si and ZnO/GaN interfaces, significantly exceed theoretical results or even the absolute upper limit called the ``radiation limit", suggesting the failure of the theory. Here, we identify that for high-transmission interfaces, the commonly assumed phonon local thermal equilibrium adjacent to the interface fails, and the measurable phonon temperatures are not their emission temperature. We hence develop a ``nonequilibrium Landauer approach" and define the unique ``dressed" and ``intrinsic" TBCs. Combining our approach even with a simple diffuse mismatch model (DMM) nearly doubles the theoretical TBCs across the Al/Si and ZnO/GaN interfaces, and the theoretical results agree with experiments for the first time. The radiation limit is also redefined and found to increase over 100\% over the original radiation limit, and it can now well bound all the experimental data.

\end{abstract}


\maketitle
\end{CJK*}

As modern electronic devices shrink in size, heat dissipation is increasingly limited by thermal resistances across interfaces~\cite{Pop2006a, Moore2014}. For example, the thermal boundary resistance (TBR) at the CNT-graphene junction is comparable to the resistance of a 200 nm long pure CNT~\cite{Shi2015, Shi2018_acs}. Hence, understanding thermal interfacial transport and designing interfaces with high thermal boundary conductance (TBC) are urgently needed for thermal management. Many such interfaces, including the highly-matched Al/Si and ZnO/GaN interfaces, have been experimentally measured ~\cite{Minnich2011a, Monachon2013, Wilson2014, Gaskins2018}. However, the TBCs cannot be explained by the widely used Landauer theory, which significantly underestimates the TBCs~\cite{Huberman1994, Hu2009, Gaskins2018, Monachon2016}. The neglect of inelastic phonon scattering in the transmission function is often considered to be responsible for the discrepancy~\cite{Hopkins2009,Paulsson2005}.  However, even the absolute theoretical upper limit called the radiation limit that assumes a unity transmission still often under-estimates TBCs~\cite{Swartz1989, Stoner1993, Lyeo2006}, suggesting the failure of the Landauer approach and the warrant of new theories~\cite{Gaskins2018}.    
 
\begin{figure}
 	\includegraphics[width=0.8\columnwidth]{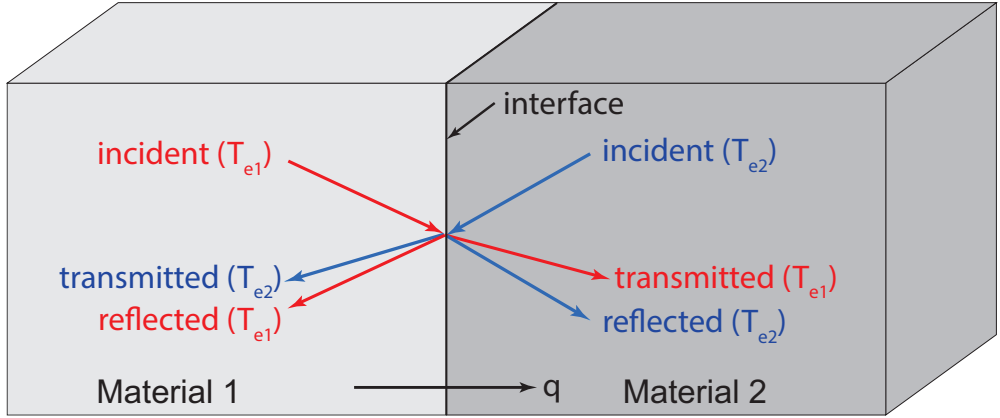}%
 	\caption{ The general structure of thermal transport considered in Landauer approach.\label{fig_structure}}
 \end{figure}

The Landauer theory, which is built on the particle description of phonons, can be examined more closely. The heat flux is the net between the two fluxes flowing in opporsite directions, while the temperature drop is between the phonon emitted temperatures $T_e$ of the two thermal reserviors~\cite{Landauer1957, Little1959, Swartz1989, Stoner1993, Jeong2012, fisher2013, Reddy2005, Duda2010, Li2012, Norris2013, Gaskins2018, Zhou2018}. Clearly, the Landauer approach is based on an important but often ignored assumption that the interface is a locally thermal-equilibrium system, where the measurable temperatures of all phonon modes are the emission temperatures. Such an assumption may be reasonable for electron interfacial transport and photon radiative transer across a gap, for which the Landauer approach was originally developed, since the temperature reservoirs can be held right adjacent to the interface or surface, and the local temperature near the interface is almost not affected by the transmitted electrons or photons. However, it is questionable for phonons, since in standard measurements of interfacial thermal conductance such as the two-bar method \cite{Rosochowska2004}, the reservoirs are placed far away from the interface. Recently, a modal nonequilibrium molecular dynamics approach~\cite{Feng2017} shows that different phonon modes can be driven into strong nonequilibrium near an interface, while a clear analytical model is not available yet. It was also pointed out in previous works by Simons~\cite{Simons1974b}, and Zeng and Chen~\cite{ZengGangChen2001} that on each side of the interface (as shown in FIG.~\ref{fig_structure}) there are three groups of phonons. Take the left side as an example, we see incident phonons traveling toward the interface with $T_{e,1}$, reflected phonons with $T_{e,1}$, and transmitted phonons from the right-side with $T_{e,2}$. 
These phonons are in strong nonequilibrium because of the highly mode-dependent reflection and transmission coefficients together with the different temperatures of reservoirs. If the transmission coefficient is not low, the measurable temperature near the interface is no longer simply the incident (or emitted) temperature $T_e$ but very different. 
A local equivalent equilibrium temperature was defined in Zeng and Chen's work~\cite{ZengGangChen2001}, and it correctly predicts a zero resistance for an imaginary interface in a pure material (the transmission coefficients $\tau_{12}$ and $\tau_{21}$ are both 1) rather than a finite resistance that would be predicted by the original Landauer formula. However, the model was a gray approach and was not used or validated on real materials interfaces which would have rich spectral transmission characteristics. Landry and McGaughey~\cite{Landry2009}, and Merabia and Termentzidis~\cite{Merabia2012} have suggested approaches to include nonequilibrium phonon distributions in molecular dynamics, inspired by Simons~\cite{Simons1974b}. However, probably due to the inherent inaccuracy in the interfacial interatomic potentials, none of these works have been compared to experiments to show that the agreement can be improved.

In this work, we develop a ``non-equilibrium Landauer approach" to capture modal phonon non-equilibrium adjacent to interfaces, and define two unique concepts, i.e., ``intrinsic" and ``dressed" interfacial conductance. Our model paired even with a simple diffuse mismatch model (DMM) is shown to significantly improve the agreement between theory and experiment. We first apply the original Landauer method to the Al/Si interface, whose TBC has been measured by several groups and consistent results have been reported~\cite{Minnich2011a, Monachon2013, Wilson2014}. However, both our calculated results and those from literature based on the original Landauer formula underestimates the TBC by over 25\%. For the ZnO/GaN interface, the underestimation is as high as 56\%~\cite{Gaskins2018}. We then propose that the underestimation comes from the local equilibrium assumption. We define the modal equivalent equilibrium temperature of the corresponding phonon mode, and develop the nonequilibrium Landauer approach. After that, we calculate the modal conductance and uniquely define the ``dressed" and ``intrinsic" interfacial thermal conductances $G_\text{dressed}$ and $G_\text{intrinsic}$. It should be noted that these two types of conductances can only be distinguished using our modal approach while not in the gray approach in Refs.~\cite{ZengGangChen2001}. The results from our approach can now well explain the experimental data for both Al/Si and ZnO/GaN interfaces.

\section{Results}
\subsection{Failure of the original Landauer approach at the Al/Si interface.}
We begin with the original Landauer approach, and the TBC at the Al/Si interface is calculated from the net heat flow rate $q$, cross-sectional area $A_{c}$ of the interface, and emitted phonon temperatures $T_{e,1}$ and $T_{e,2}$ as shown in the following equation:
\begin{equation}
G _{original}= \frac{q}{A_{c} \Delta T} = \frac{\sum\limits_{\nu}\int_{0}^{+\infty}\frac{1}{4} D_\lambda v_\lambda \tau_\lambda \hbar\omega(f_1-f_2) d\omega}{T_{e,1}-T_{e,2}} = \sum\limits_{\nu}\int_{0}^{+\infty}\frac{1}{4} D_\lambda v_\lambda \tau_\lambda \hbar\omega \frac{df}{dT} d\omega,
\label{eqR}
\end{equation}
where $\lambda$ is short for phonon mode $(\omega,\nu)$ with $\omega$ and $\nu$ representing the phonon frequency and dispersion branch, respectively. The $\nu$-sum $\sum\limits_{\nu}$ runs over all the incident phonon branches, $\hbar$ is the reduced Planck constant, $\omega$ is the phonon angular frequency, $D$ is the phonon density of states, $v$ is the modal phonon group velocity, $\tau$ is the modal transmission coefficient at the interface from one material to the other, $f_1$ and $f_2$ are short for $f(\omega,T_{e,1})$ and $f(\omega,T_{e,2})$, which are the carrier statistics (the Bose-Einstein distribution function is used for phonons) at $T_{e,1}$ and $T_{e,2}$ (emitted phonon temperatures from two reservoirs) with a certain frequency $\omega$, and $df/dT$ includes the assumption that $T_{e,1}-T_{e,2}$ equals to $\Delta T$.

The net heat flow rate $q$ in Eq.(\ref{eqR}) as shown in FIG.~\ref{fig_structure} can also be expressed in the following form~\cite{fisher2013}:
\begin{equation}
q=\sum\limits_{\nu}\int_{0}^{+\infty} \frac{1}{2\pi} M_\lambda \tau_\lambda \hbar\omega (f_1-f_2) d\omega,
\label{eq1}
\end{equation}
where $M_\lambda$ is the phonon number of modes, and the transmission function $\tau$ is calculated from the diffuse mismatch model (DMM). The DMM is used here instead of acoustic mismatch model (AMM)~\cite{Swartz1989} which is another commonly used model, because we intend to compare with experiments. The AMM usually works well for smooth interfaces at low temperature, while there are different kinds of defects at interfaces in experiments with which the DMM usually shows better agreements~\cite{Costescu2003, shin_roles_2010}. The detailed derivation of the Landauer formula can be found in the Supplementary Information. The expression with number of modes is introduced here to simplify the transmission function later. It should be noted that only the phonon properties of material 1 is needed here because of the constraint from detailed balance, that the net heat flow rate is 0 when $T_{e,1}=T_{e,2}$, and the information of material 2 is included in the transmission coefficient $\tau$. The related proof can be found in Supplementary Information. Moreover, the Eq.(\ref{eq1}) is for interfaces with transmission coefficients independent of phonon incident angle, or at least one of material 1 and 2 is one dimensional (1D). This approximation is valid in this work, as the transmission from DMM does not depend on the angle of incidence~\cite{Swartz1989}.

The phonon properties of both Al and Si are obtained from ab initio calculations within the framework of density-functional theory (DFT), as implemented in the Vienna Ab initio Simulation Package (VASP)~\cite{Kresse1993,Kresse1996}, the dispersion curves are shown in FIG.~\ref{fig_AlSi_disp} and the details of calculation can be found in Supplementary Information . It is a highly matched interface due to the small atomic mass mismatch between Al and Si. 

\begin{figure}[t]
	\includegraphics[width=0.6\columnwidth]{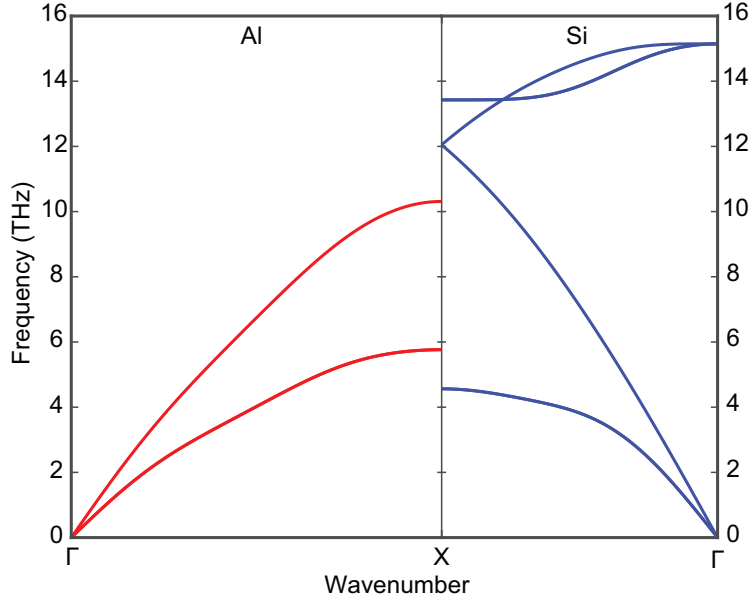}%
	\caption{ The phonon dispersion relations of Al and Si calculated from first-principles method.\label{fig_AlSi_disp}}
\end{figure}

\begin{figure}[b]
	\includegraphics[width=0.8\columnwidth]{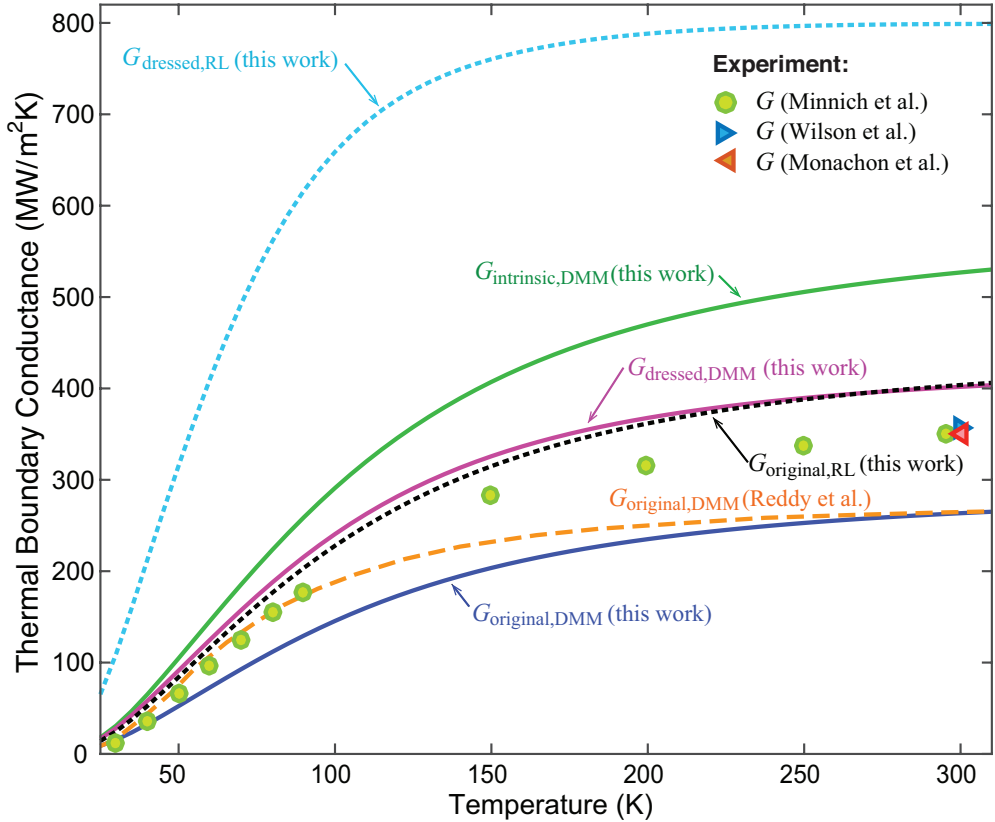}%
	\caption{ The comparison of TBC across Al/Si interface between different Landauer approaches and Experiment. The previous calculations from Reddy~\cite{Reddy2005} and measurements from Minnich~\cite{Minnich2011a}, Wilson~\cite{Wilson2014}, and Monachon~\cite{Monachon2013} are also included.
	\label{fig_G_AlSi}}
\end{figure}

The calculated spectral transmission coefficients $\tau$ from DMM can be found in the Supplementary Information. 
The temperature-dependent conductance results from the original Landauer approach are then compared to experimental data ~\cite{Minnich2011a, Monachon2013, Wilson2014} as well as existing calculations from \cite{Reddy2005} as shown in FIG.~\ref{fig_G_AlSi}. The original radiation limit $G_\text{original,RL}$ is also shown. Clearly, both our results and those from Ref. \cite{Reddy2005} based on the original Landauer approach underestimate the TBC.

\subsection{Nonequilibrium Landauer approach at the Al/Si interface}
\begin{figure}
	\includegraphics[width=1.0\columnwidth]{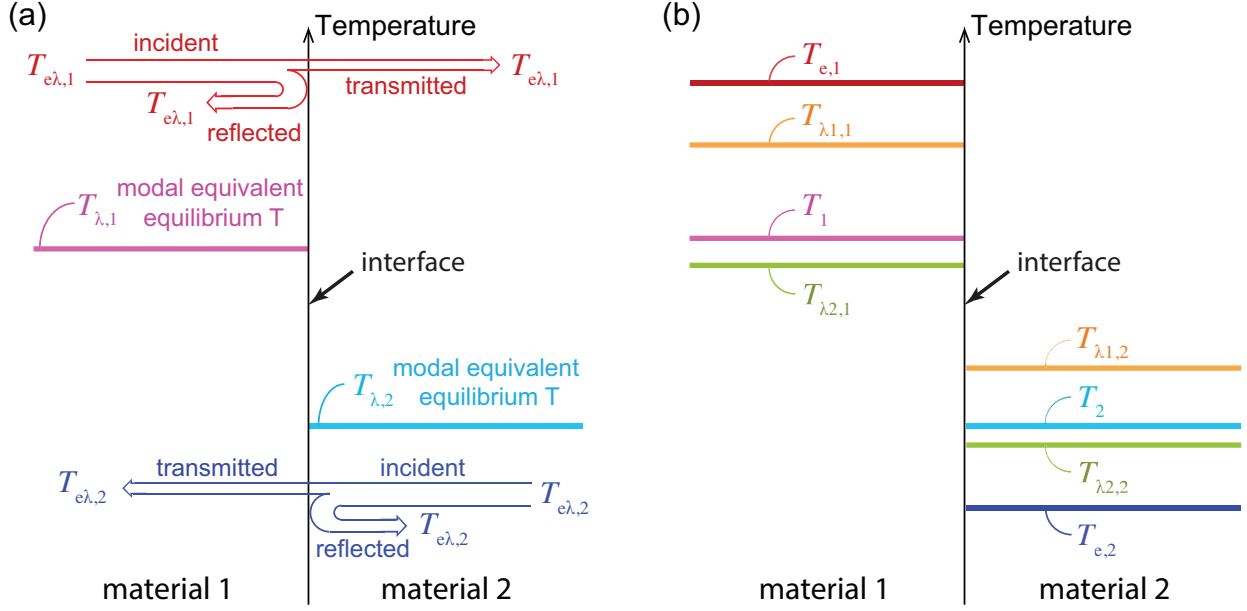}%
	\caption{ (a) The schematic of modal emitted temperature $T_{e\lambda}$ and modal equivalent equilibrium temperature $T_\lambda$. (b) The relation between the local temperature $T$ near the interface and the modal equivalent equilibrium temperature $T_\lambda$ of two different phonon modes. \label{fig_MEET}}
\end{figure}

We suspect that the underestimation comes from the local nonequilibrium adjacent to the interface, and hence we closely examine it at the interface as shown in FIG.~\ref{fig_MEET}(a). At the interface, for each phonon mode there are incident phonons, reflected phonons with the modal reflection coefficient (1-$\tau_\lambda$), and transmitted phonons from the other side. 
These phonons are from reservoirs with different temperatures and a modal equivalent equilibrium temperature $T_\lambda$ is defined as shown in FIG.~\ref{fig_MEET}(a). 
For highly matched interfaces where the transmission coefficients are high, the measurable temperature can now be significantly affected by the transmitted phonons and is no longer the same as the emitted phonon temperature $T_e$. Clearly, the original Landauer approach misses this nonequilibrium physics that experiment captures. The modal equivalent equilibrium temperatures $T_{\lambda,1}$ and $T_{\lambda,2}$ are calculated as:
\begin{equation}
T_{\lambda, 1} = T_{e, 1} -\tau_{12,\lambda}(T_{e, 1} -T_{e, 2})/2 ,
\end{equation}
\begin{equation}
T_{\lambda, 2} = T_{e, 2} +\tau_{21,\lambda}(T_{e, 1} -T_{e, 2})/2,
\end{equation}
and the detailed derivation process of $T_\lambda$ is in the Methods section.
According to $T_\lambda$, a modal conductance $G_\lambda$ can be defined as:
\begin{equation}
G_\lambda = \frac {M_\lambda \tau_\lambda \hbar\omega (f_1-f_2) d\omega} {T_{\lambda, 1}-T_{\lambda, 2}}.
\end{equation}
Clearly $G_\lambda$ will be higher than that from the original Landauer approach since the temperature difference in the denominator is decreased compared to that in the original Landauer approach.

\subsection{Dressed and intrinsic TBC across the Al/Si interface}

To calculate the total TBC, now we need to determine the overall equivalent equilibrium temperatures by including all phonon modes. This is actually a very difficult task. One can easily assume that all phonon modes have the same $T_{e \lambda}$ adjacent to the interface. However, it was pointed out in previous modal NEMD simulations~\cite{Feng2017} and multi-temperature model ~\cite{Lu2019} that, in a typical experimental interface between two semi-infinite solids, the thermostats are placed far enough from the interface, so different phonon modes arrive at the interface at different emitted temperatures although they leave the thermostat with the same temperature. On the other hand, different phonon modes will not share the same $T_\lambda$ either, similarly due to the coupling between interfacial transport and bulk transport in the leads \cite{Lu2019}. A rigorous approach will calculate TBC by using $G_\lambda$ in the multi-temperature model \cite{Lu2019} that covers both the interface and two leads. However, here without going into too much detail in the two leads, we can assume the same $T_{e \lambda}$ and $T_\lambda$ as two limiting cases, to derive the TBC respectively.    

We first determine the TBC as we assume that all phonon modes have the same emitted temperatures $T_{e,1}$ and $T_{e,2}$ on the two sides respectively, shown in FIG.~\ref{fig_MEET}(b). The overall equivalent equilibrium temperatures $T_1$ and $T_2$, which are the local temperature on each side of the interface as shown in FIG.~\ref{fig_MEET}(b), can be easily written in terms of $T_\lambda$:
\begin{equation}
T_1 =\frac{\sum\limits_{\nu}\int_{0}^{+\infty} w_{\lambda, 1} T_{\lambda, 1} d\omega}{ \sum\limits_{\nu}\int_{0}^{+\infty} w_{\lambda, 1} d\omega},
\label{eqT1}
\end{equation}
\begin{equation}
T_2 =\frac{\sum\limits_{\nu}\int_{0}^{+\infty} w_{\lambda, 2} T_{\lambda, 2} d\omega}{ \sum\limits_{\nu}\int_{0}^{+\infty} w_{\lambda, 2} d\omega},
\label{eqT2}
\end{equation}
where $w_{\lambda, 1}$ and $w_{\lambda, 2}$ are the modal equivalent equilibrium temperature weights of the corresponding phonon modes in material 1 and 2, and equal to the total incident energy to the interface of a certain mode $\lambda$, and the expressions are defined in the Methods section, and the TBC can be calculated as:
\begin{equation}
\begin{split}
G_\text{dressed} & = \frac{q}{T_1-T_2}. 
\label{G_dressed}
\end{split}
\end{equation}
We call the calculated conductance dressed TBC because the modal $G_\lambda$ is dressed by the overall temperature jump $T_1 - T_2$ rather than their respective $T_{\lambda,1}-T_{\lambda,2}$, hence the total $G_\text{dressed}$ is not a direct summation of $G_\lambda$). Moreover, we define a dressed radiation limit $G_\text{dressed,RL}$ as shown in FIG.~\ref{fig_G_AlSi}. Transmission function is the only difference between $G_\text{dressed,RL}$ and $G_\text{dressed}$, and the radiation limit transmission coefficients can be found in the Methods section. 

We then derive the TBC by assuming that the modal equivalent equilibrium temperature $T_\lambda$ adjacent to the interface for all the modes are the same. This assumption would generally require the phonons to have different modal emitted temperatures $T_{e \lambda}$ (unless a constant transmission coefficient which does not change with frequency or even a unity transmission for all phonon modes which recovers the imaginary interface case). Now all phonons are actually in local thermal equilibrium which gives what we define as the intrinsic conductance $G_\text{intrinsic}$, as:
\begin{equation}
G_\text{intrinsic} = \frac{q}{\Delta T} = \frac{\sum\limits_{\lambda} q_\lambda}{\Delta T_{\lambda}}  = \sum\limits_{\lambda} \frac{q_\lambda}{\Delta T_{\lambda}} = \sum\limits_{\lambda} G_\lambda.
\end{equation}
We call it $G_\text{intrinsic}$ because it is a simple summation of the modal conductance $G_\lambda$ due to the fact that all the phonon modes are in local equilibrium, unlike $G_\text{dressed}$. The intrinsic radiation limit is difficult to define though. The nominal intrinsic radiation limit would be infinity since all phonon modes having unity transmission will have infinity modal conductances. However, in real interfaces, all the modes with unity transmission coefficient will have a zero $\Delta T_\lambda$ but the rest modes won't, hence the local thermal equilibrium will no longer hold. For this reason, we do not define an intrinsic radiation limit. We should note, however, for imaginary interfaces all the modes will have unity transmission and zero $\Delta T_\lambda$, hence the local equilibrium is valid again and the interfacial conductance is indeed infinity.

The dressed and intrinsic conductances of Al/Si interface are shown in FIG.~\ref{fig_G_AlSi}. Both are significantly larger than that from the original Landauer approach and even the original radiation limit, and can now bound the experimental data at the entire temperature range. These two conductances can only be distinguished using our modal and nonequilibrium approach while not in the gray approach where all phonon modes have the same $\tau$, $T_e$, $T_1$, and $T_2$ at the same time. Our modal nonequilibrium approach brings quantitative agreement with experiment in reach. We can also see that $G_\text{intrinsic}$ is generally higher than $G_\text{dressed}$, and this is due to the modal feature captured in our approach. For $G_\text{dressed}$, we can see from FIG.~\ref{fig_MEET}(a) that the transmissible modes (below the cutoff frequency) will have a lower $\Delta T_\lambda$ than that for modes with zero transmission coefficient (above the cutoff). As a result, these transmissible modes under-contribute to the interfacial conductance as compared to the $G_\text{intrinsic}$ case where their $\Delta T_\lambda$ are raised to be the same as those zero transmission modes. The spectral and accumulated TBC plots of $G_\text{original}$, $G_\text{dressed}$ and $G_\text{intrinsic}$ at room temperature can be found in the Supplementary Information. Besides, we can note that the dressed radiation limit $G_\text{dressed,RL}$ shows over 100\% increase over the original radiation limit $G_\text{original,RL}$,  and well bounds the experimental data.

\subsection{TBC at the ZnO/GaN interface.}

\begin{figure}
	\includegraphics[width=0.8\columnwidth]{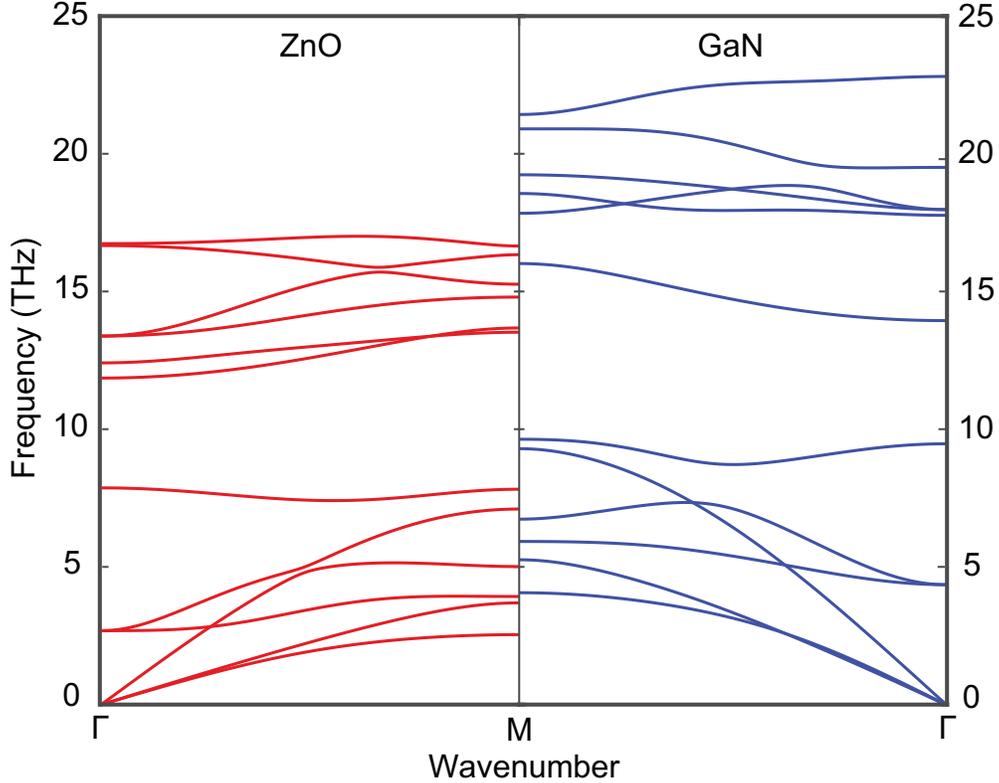}%
	\caption{ The phonon dispersion relations of ZnO and GaN calculated from first principles.\label{fig_ZnOGaN_disp}}
\end{figure}

We also apply our nonequilibrium Landauer approach to the ZnO/GaN interface to see if our approach works well. As mentioned earlier, the original Landauer approach was found to under-predict the TBC of ZnO/GaN interface by nearly a factor of two ~\cite{Gaskins2018}. The phonon dispersion relations of ZnO and GaN are calculated from first principles as shown in FIG.~\ref{fig_ZnOGaN_disp}. Then the original and nonequilibrium Landauer approaches, and the radiation limit calculations are performed. The transmission coefficients from DMM can be found in Supplementary Information. As shown clearly in FIG.~\ref{fig_G_ZnOGaN}, the experimental results are much higher than the original Landauer results based on DMM or atomistic Green's function (AGF) method ~\cite{Gaskins2018}. In fact, the measurements even exceed the original radiation limit $G_\text{original,RL}$. With our nonequilibrium Landauer approach, TBC results from both $G_\text{dressed}$ and $G_\text{intrinsic}$ are much higher than $G_\text{original}$ and agree with experimental data much better. At 300 K and above, the experimental value still exceeds our calculations, probably due to inelastic phonon scattering. But the inelastic scattering is not very strong as compared to the non-equilibrium effect. This is understandable for the highly-matched interface. On the other hand, it can be seen that $G_\text{dressed,RL}$ can now bound the experimental data. This has very important implications that the Landauer may still be a valid theory for such interfaces if used in the correct way, despite previous suggestions that new theories are needed. The spectral and accumulated TBC plots of $G_\text{original}$, $G_\text{dressed}$ and $G_\text{intrinsic}$ at room temperature can be found in the Supplementary Information.

\begin{figure}
	\includegraphics[width=0.8\columnwidth]{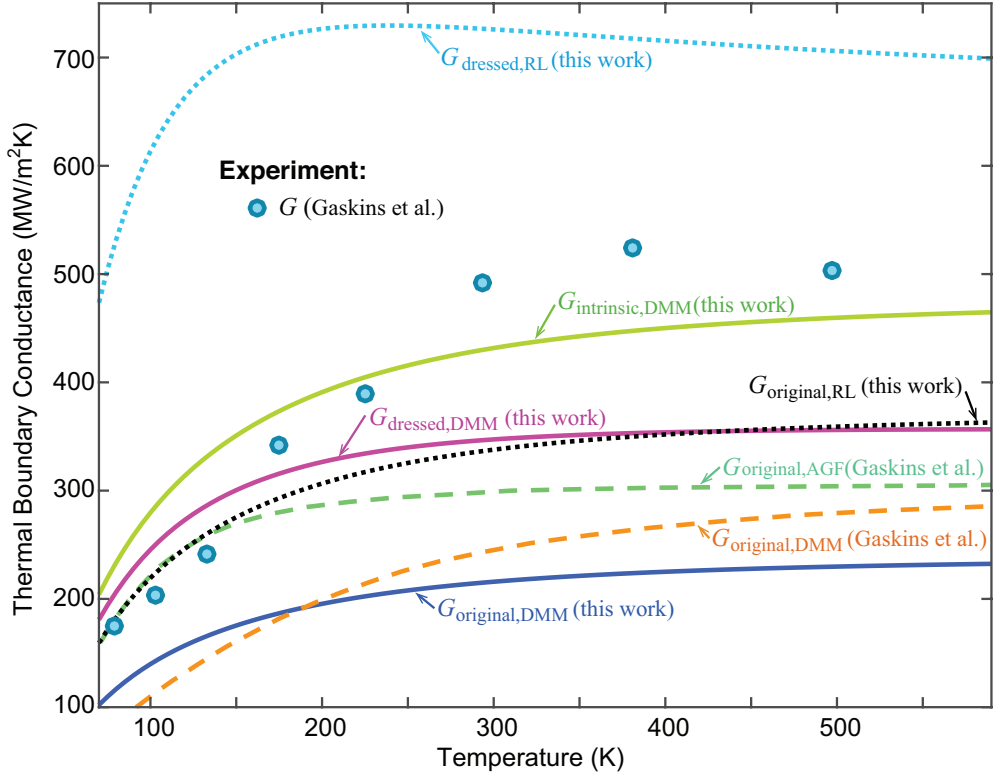}%
	\caption{ The comparison of TBCs across the ZnO/GaN interface from experiments and different Landauer approaches. The modeling and experimental results from Ref.~\cite{Gaskins2018} are also included. \label{fig_G_ZnOGaN}}
\end{figure}

\section{Discussion}

From the plot of TBC at ZnO/GaN interface in FIG.~\ref{fig_G_ZnOGaN}, it can be seen that $G_\text{dressed}$ and $G_\text{dressed,RL}$ start to decrease at high temperature. This is due to the excitation of more high-frequency modes, leading $T_1-T_2$ closer to $T_{e,1}-T_{e,2}$ at higher temperature. Because of the different cut-off frequency of ZnO and GaN, the high frequency phonons with emitted temperature from GaN cannot transmit to ZnO in our elastic Landauer approach. As the temperature increases, the percentage of these zero-transmission high frequency phonons will also increase, leading the local temperature near the interface $T$ to approach the emitted temperature $T_e$. As a result, the TBC will decrease with increasing temperature. However, usually we cannot observe this behavior in experiments because the inelastic scattering, which is not included in our model, will also be stronger at higher temperature.

It is a natural question when we need to consider the non-equilibrium effect. We have seen that for the highly matched interface, the effect is significant. For other typical disimilar interfaces like Si-Ge, we expect the effect to be weaker but non-negligible. For highly-mismatched interfaces such as diamond-Pb, the transmission coefficients are generally low and the non-equilibrium effect is expected to be unimportant. However, we should note that significant under-prediction from the Landauer approach was also found for such interfaces~\cite{Lyeo2006, Huberman1994}. This is due to that the inelastic scattering becomes relatively more important for such highly-mismatched interfaces~\cite{Wilson2014,Hopkins2009,Paulsson2005}. Our work offers very interesting insights that although both highly-matched and highly-mismatched interfaces were under-estimated, they are of very different nature. 
When the inelastic phonon scattering is included for highly-mismatched interfaces,  the modal phonon transmission will increase, especially for those above the cutoff, hence the non-equilibrium effect presented here will also become more important. 
Moreover, while evaluating the radiation limit for highly mismatched interfaces, the non-equilibrium effect will be very important again because of the assumed unity transmission coefficient.

\begin{figure}
	\includegraphics[width=0.7\columnwidth]{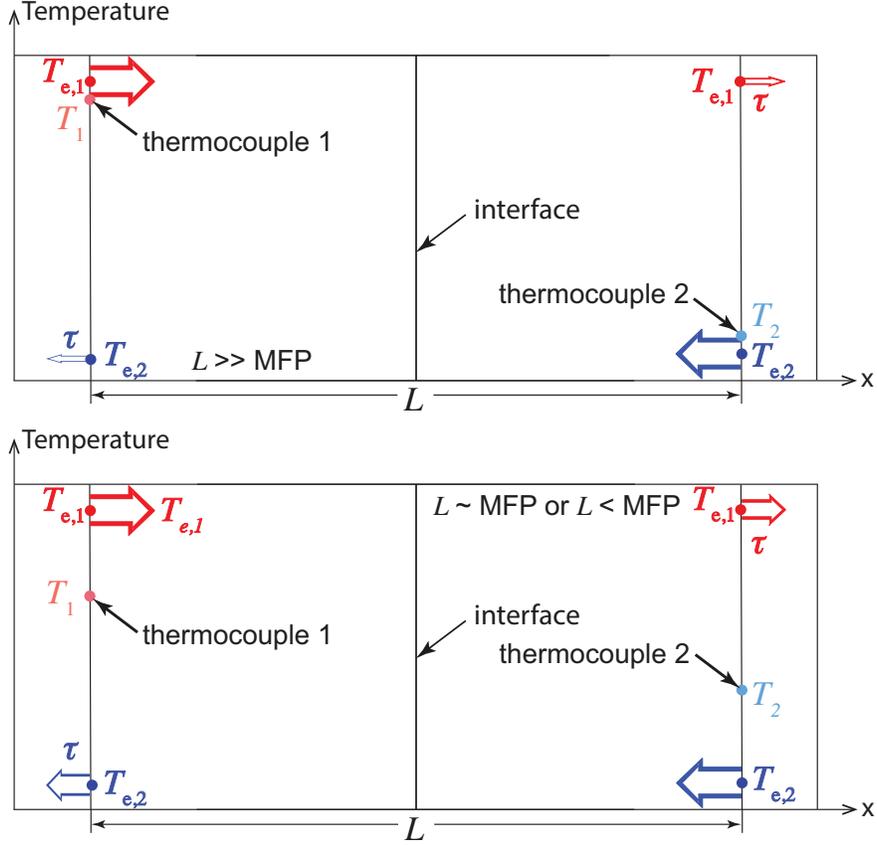}%
	\caption{ The schematic of ``measurable" temperatures at contacts of device with different lengths.\label{fig_AGF}}
\end{figure}

We also note that our nonequilibrium Landauer approach can be used with DMM, AMM (acoustic mismatch model), AGF (the atomic Green's function), and any other methods of calculating the transmission function. A special note is given here to AGF.  Different from DMM or AMM, the AGF method includes two short leads in the deviance region besides the interface.  A four-probe method has been proposed to subtract the resistances of the leads in order to isolate the interfacial resistance~\cite{Tian2012}.  However, the interfacial conductance utilizing four-probe was still often under-predicted compared to experiments~\cite{Gaskins2018}.  We believe the temperature correction presented in this work should also be considered.  As shown in FIG.~\ref{fig_AGF}, if the device length is much larger than the phonon mean free path (MFP), the transmission coefficient of phonons starting from one contact and reaching the other contact is very small, hence the ``measurable" temperatures (such as by a thermocouple as shown in FIG.~\ref{fig_AGF}) are close to the emission temperatures, and no correction is needed. However, if the device length is comparable to or smaller than the phonon MFP, as is the case of AGF, the transmission coefficient of phonons starting from one contact and reaching the other contact will be large, hence the ``measurable" temperature will deviate much from the emission temperatures. If the non-equilibrium effect is not included, both the conductances of the device and leads will be under-estimated. Although the four-probe approach can indeed recover the infinity conductance for an imaginary interface due to the cancellation of all the under-estimations, it alone will under-estimate the conductance of a real interface with finite conductance. Using the non-equilibrium Landauer approach in this work is expected to scale up the interfacial conductance.

To summarize, in this work we show that the effect of local thermal non-equilibrium is important for high-matched interfaces. We have developed a ``non-equilibrium Landauer approach" to capture such effects, and defined dressed and intrinsic conductances to include the phonon modal characteristics. Our approach coupled with the simple DMM on Al/Si and ZnO/GaN interfaces yield results in much better agreement with experiments than the original Landauer with both DMM and the sophisticated AGF methods. We have also re-defined the radiation limit and shown that the experimental TBCs do not actually exceed the radiation limit. Our work sheds important new insights in interpreting experimental data for highly-matched thermal interfaces.

\section{Methods}
\subsection{Original Landauer approach}
In the calculation of heat flow rate in Eq.(\ref{eq1}), the number of modes for the interface between three dimensional (3D) materials is:
\begin{equation}
M_{3D}=\frac{A_c k_\lambda^2}{4\pi},
\end{equation}
where $k_\lambda$ is the wave number at a certain frequency $\omega$, and an isotropic assumption is included in the expression.

The transmission coefficient of DMM is defined as~\cite{Swartz1989, fisher2013}:
\begin{equation}
\tau_{12}(\omega)=1-\tau_{21}(\omega)=\frac{\sum\limits_{\nu}M_2(\omega)}{\sum\limits_{\nu}M_1(\omega)+\sum\limits_{\nu}M_2(\omega)}.
\label{eqDMM}
\end{equation}
The detailed balance of Landauer formula with transmission function from DMM is automatically satisfied because of the DMM definition, and a proof can be found in the Supplementary Information. Because from DMM model, all the incident modes lose their memory during transmission process, the transmission coefficients of DMM only depend on the frequency but not polarization unlike AMM. Both DMM and AMM assume elastic transmission process without the change of phonon frequency.
With all the information, TBC $G_\text{original}$ from the original Landauer approach at the Al/Si interface can now be calculated as:
\begin{equation}
G_\text{original}=\frac{\sum\limits_{\nu}\int_{0}^{+\infty}\frac{1}{8\pi^2}{k_\lambda^2\tau_\lambda\hbar\omega (f_1-f_2) }d\omega}{T_{e,1}-T_{e,2}}.
\end{equation}

\subsection{Radiation limit}
The equation to calculate the original radiation limit $G_\text{original,RL}$ is:
\begin{equation}
G_\text{original,RL}=\frac{\sum\limits_{\nu}\int_{0}^{+\infty}\frac{1}{8\pi^2}{k_\lambda^2\tau_{\text{RL},\lambda} \hbar\omega (f_1-f_2)}d\omega}{T_{e,1}-T_{e,2}},
\end{equation}
and the radiation limit transmission $\tau_{\text{RL}}$ at a certain frequency $\omega$ is:
\begin{equation}
\tau_{\text{RL},12}(\omega) = 
  \begin{cases} 
  0 & \text{ if } \omega> \text{min} \{\omega_\text{cutoff,1},\omega_\text{cutoff,2}\} \\ 1 & \text{ if } \sum\limits_{\nu}M_1(\omega) \leq \sum\limits_{\nu}M_2(\omega) \\
   \frac{\sum\limits_{\nu}M_2(\omega)}{\sum\limits_{\nu}M_1(\omega)} & \text{ if } \sum\limits_{\nu}M_1(\omega)>\sum\limits_{\nu}M_2(\omega)
  \end{cases}
  \label{eq_RLtran1}
\end{equation}
\begin{equation}
\tau_{\text{RL},21}(\omega) = 
  \begin{cases} 
  0 & \text{ if } \omega>\text{min}\{\omega_\text{cutoff,1},\omega_\text{cutoff,2}\} \\ \frac{\sum\limits_{\nu}M_1(\omega)}{\sum\limits_{\nu}M_2(\omega)} & \text{ if } \sum\limits_{\nu}M_1(\omega) \leq \sum\limits_{\nu}M_2(\omega) \\
  1 & \text{ if } \sum\limits_{\nu}M_1(\omega)>\sum\limits_{\nu}M_2(\omega)
  \end{cases}
  \label{eq_RLtran2}
\end{equation}
With the transmission defined like this, the transmission from either side of the interface will not exceed 1 and the detailed balance is fulfilled. 

\subsection{Nonequilibrium Landauer approach}
Because there are incident, reflected and transmitted phonons at the interface, the local equivalent equilibrium temperatures $T_1$ and $T_2$ can be expressed as:
\begin{equation}
T_1 =\frac{\parbox{4in}{$\sum\limits_{\nu}\int_{0}^{+\infty} T_{e, 1} \hbar\omega D_{\lambda,1} v_{\lambda,1} f_1 d\omega $ \\ 
\hspace*{1.5cm}$ +\sum\limits_{\nu}\int_{0}^{+\infty} T_{e,1} \hbar\omega D_{\lambda,1} v_{\lambda,1} f_1 (1-\tau_{12,\lambda}) d\omega $ \\ 
\hspace*{3.5cm}$  +\sum\limits_{\nu}\int_{0}^{+\infty} T_{e,2} \hbar \omega D_{\lambda,2} v_{\lambda,2} f_2 \tau_{21,\lambda} d\omega$}}{\parbox{4in}{$\sum\limits_{\nu}\int_{0}^{+\infty} \hbar\omega D_{\lambda,1} v_{\lambda,1} f_1 d\omega $ \\ 
\hspace*{1.5cm}$  +\sum\limits_{\nu}\int_{0}^{+\infty} \hbar\omega D_{\lambda,1} v_{\lambda,1} f_1 (1-\tau_{12,\lambda}) d\omega $ \\ 
\hspace*{3.5cm}$  +\sum\limits_{\nu}\int_{0}^{+\infty} \hbar \omega D_{\lambda,2} v_{\lambda,2} f_2 \tau_{21,\lambda} d\omega$}},
\label{T1}
\end{equation}
\begin{equation}
T_2 =\frac{\parbox{4in}{$\sum\limits_{\nu}\int_{0}^{+\infty} T_{e, 2} \hbar\omega D_{\lambda,2} v_{\lambda,2} f_2 d\omega  $ \\ 
\hspace*{1.5cm}$ +\sum\limits_{\nu}\int_{0}^{+\infty} T_{e, 2} \hbar\omega D_{\lambda,2} v_{\lambda,2} f_2 (1-\tau_{21,\lambda}) d\omega $ \\ 
\hspace*{3.5cm}$   +\sum\limits_{\nu}\int_{0}^{+\infty} T_{e,1} \hbar \omega D_{\lambda,1} v_{\lambda,1} f_1 \tau_{12,\lambda} d\omega$}}{\parbox{4in}{$\sum\limits_{\nu}\int_{0}^{+\infty}  \hbar\omega D_{\lambda,2} v_{\lambda,2} f_2 d\omega $ \\ 
\hspace*{1.5cm}$  + \sum\limits_{\nu}\int_{0}^{+\infty} \hbar\omega D_{\lambda,2} v_{\lambda,2} f_2 (1-\tau_{21,\lambda}) d\omega $ \\ 
\hspace*{3.5cm}$   + \sum\limits_{\nu}\int_{0}^{+\infty} \hbar \omega D_{\lambda,1} v_{\lambda,1} f_1 \tau_{12,\lambda} d\omega$}}.
\label{T2}
\end{equation}
We would like to simplify the expression of $T_1$ and $T_2$ in Eq.(\ref{T1}) and (\ref{T2}),
If the difference between $T_{e,1}$ and $T_{e,2}$ is small, the $f_1$ and $f_2$ will be close to each other (details can be found in the Supplementary Information), and an $f_0$ is used to approximate them. 
In addition, from the detailed balance at different temperature, we have:
\begin{equation}
\sum\limits_{\nu} D_{\lambda,1} v_{\lambda,1} \tau_{12,\lambda}=\sum\limits_{\nu} D_{\lambda,2} v_{\lambda,2} \tau_{21,\lambda}.
\end{equation}
With the Bose-Einstein distribution approximation and the detailed balance relation, $T_1$ and $T_2$ can be simplified as:
\begin{equation}
T_1 =\frac{\sum\limits_{\nu}\int_{0}^{+\infty} \hbar\omega D_{\lambda,1} v_{\lambda,1} f_0 [T_{e, 1} (2-\tau_{12,\lambda}) +T_{e,2}\tau_{12,\lambda}]d\omega}{ 2 \sum\limits_{\nu}\int_{0}^{+\infty} \hbar \omega D_{\lambda,1} v_{\lambda,1} f_0d\omega},
\end{equation}
\begin{equation}
T_2 =\frac{\sum\limits_{\nu}\int_{0}^{+\infty} \hbar\omega D_{\lambda,2} v_{\lambda,2} f_0 [T_{e, 2} (2-\tau_{21,\lambda}) +T_{e,1}\tau_{21,\lambda}]d\omega}{ 2 \sum\limits_{\nu}\int_{0}^{+\infty} \hbar \omega D_{\lambda,2} v_{\lambda,2} f_0d\omega}.
\end{equation}
From the expression of $T_1$ and $T_2$, the modal equivalent equilibrium temperature $T_{\lambda, 1}$ and $T_{\lambda, 2}$ in material 1 and 2 are defined as:
\begin{equation}
T_{\lambda, 1} =T_{e, 1} (1-\frac{\tau_{12,\lambda}}{2}) +T_{e,2}\frac{\tau_{12,\lambda}}{2} = T_{e, 1} -\tau_{12,\lambda}(T_{e, 1} -T_{e, 2})/2 ,
\end{equation}
\begin{equation}
T_{\lambda, 2} =T_{e, 2} (1-\frac{\tau_{21,\lambda}}{2}) +T_{e,1}\frac{\tau_{21,\lambda}}{2} = T_{e, 2} +\tau_{21,\lambda}(T_{e, 1} -T_{e, 2})/2,
\end{equation}
and the schematic of modal equivalent equilibrium temperature $T_{\lambda}$ of two phonon modes $\lambda_1$ and $\lambda_2$ are shown in FIG.~\ref{fig_MEET}(b). 
Because $T_\lambda$ is only for a single phonon mode on one side of the interface, $T_\lambda$ of the same frequency across the interface is used as the $T_\lambda$ on the other side in the calculation of $G_\lambda$. 

With the definition of modal equivalent equilibrium temperature, the local temperature near the interface $T_1$ and $T_2$ can now be expressed as a function of $T_\lambda$:
\begin{equation}
T_1 =\frac{\sum\limits_{\nu}\int_{0}^{+\infty} \hbar\omega D_{\lambda,1} v_{\lambda,1} f_0 T_{\lambda, 1} d\omega}{ \sum\limits_{\nu}\int_{0}^{+\infty} \hbar \omega D_{\lambda,1} v_{\lambda,1} f_0d\omega},
\end{equation}
\begin{equation}
T_2 =\frac{\sum\limits_{\nu}\int_{0}^{+\infty} \hbar\omega D_{\lambda,2} v_{\lambda,2} f_0 T_{\lambda, 2} d\omega}{ \sum\limits_{\nu}\int_{0}^{+\infty} \hbar \omega D_{\lambda,2} v_{\lambda,2} f_0d\omega}.
\end{equation}
The simplified forms are Eq.~\ref{eqT1} and~\ref{eqT2}, and the weights of $T_\lambda$ are:
\begin{equation}
w_{\lambda, 1} =\hbar\omega D_{\lambda,1} v_{\lambda,1} f_0,
\end{equation}
\begin{equation}
w_{\lambda, 2} =\hbar\omega D_{\lambda,2} v_{\lambda,2} f_0.
\end{equation}

\begin{acknowledgments}

The authors would like to acknowledge the support by the Air Force Office of Scientific Research (AFOSR) MURI grant~(FA9550-12-1-0037).

\end{acknowledgments}
\bibliography{reference}

\end{document}